\documentclass[aps,prl,superscriptaddress,twocolumn,tightenlines,showpacs,floatfix,amsmath,amssymb,pdftex,usenames,dvipsnames]{revtex4-1}

\usepackage{array}
\usepackage{amsmath}
\usepackage{amssymb}
\usepackage{bm}
\usepackage{bbm}
\usepackage{float}
\usepackage{tikz} 
\usepackage{graphicx}
\usepackage{multirow}
\usepackage{booktabs}
\usepackage[normalem]{ulem}
\usepackage{float}
\usepackage{mathtools}
\usepackage{hyperref}

\usepackage[caption=false]{subfig}

\newcommand{\mean}[1]{\langle #1\rangle}
\newcommand{\mb}[1]{\mathbf{#1}}
\newcommand{\mbb}[1]{\mathbb{#1}}
\newcommand{\bds}[1]{\boldsymbol{#1}}
\newcommand{\mr}[1]{\mathrm{#1}}

\def\multiset#1#2{\ensuremath{\left(\!\!\binom{#1}{#2}\!\!\right)}}

\DeclareMathOperator*{\sgn}{sgn}
\DeclareMathOperator*{\Tr}{Tr}

\newcolumntype{C}{>{$}c<{$}}
\AtBeginDocument{
\heavyrulewidth=.08em
\lightrulewidth=.05em
\cmidrulewidth=.03em
\belowrulesep=.65ex
\belowbottomsep=0pt
\aboverulesep=.4ex
\abovetopsep=0pt
\cmidrulesep=\doublerulesep
\cmidrulekern=.5em
\defaultaddspace=.5em
}

\begin{document}

\title{Probing hidden spin order with interpretable machine learning}
\date{\today}

\author{Jonas Greitemann}
\affiliation{Arnold Sommerfeld Center for Theoretical Physics,
Munich Center for Quantum Science and Technology,
University of Munich, Theresienstrasse 37, 80333 München, Germany}

\author{Ke Liu}
\email{ke.liu@lmu.de}
\affiliation{Arnold Sommerfeld Center for Theoretical Physics,
Munich Center for Quantum Science and Technology,
University of Munich, Theresienstrasse 37, 80333 München, Germany}

\author{Lode Pollet}
\affiliation{Arnold Sommerfeld Center for Theoretical Physics,
Munich Center for Quantum Science and Technology,
University of Munich, Theresienstrasse 37, 80333 München, Germany}

\begin{abstract}
The search of unconventional magnetic and nonmagnetic states is a major topic in the study of frustrated magnetism.
Canonical examples of those states include various spin liquids and spin nematics.
However, discerning their existence and the correct characterization is usually challenging.
Here we introduce a machine-learning protocol that can identify general nematic order and their order parameter from seemingly featureless spin configurations, thus providing comprehensive insight on the presence or absence of hidden orders.
We demonstrate the capabilities of our method by extracting the analytical form of nematic order parameter tensors up to rank $6$.
This may prove useful in the search for novel spin states and for ruling out spurious spin liquid candidates.
\end{abstract}

\maketitle
The statistical learning of phases is nowadays an active field of research~\cite{Arsenault14, Wang16, Ohtsuki16, CarrasquillaMelko17, Nieuwenburg17, Chng17, DengDasSarma17, Zhang17, Schindler17, BroeckerTrebst17, Wetzel17, Hu17, ZhangZhai18, GlasserCirac18, Cai18,LiuYH18, Beach18}.
Despite the enormous recent progress, learning or classifying intricate phases in many-body systems remains a daunting task.
Many recent algorithmic advances are tried and tested in only the simplest of models, and their applicability to more complex situations remains an open question.
The ability to interpret results to gain physical insight has been identified as one of the key challenges in the application of machine learning techniques to the domain of physics. Still, recent approaches struggle and this is only exacerbated when going beyond those simple models.
However, those situations can also be arenas for machine learning methods to demonstrate their features and prove their worth, in comparison to---or complementary to---traditional methods.

One such arena may be found in frustrated spin and spin-orbital-coupled systems~\cite{BookLacroix}.
These systems have rich phase diagrams, supporting various spin nematic (multipolar ordered)~\cite{Chubukov90, MoessnerChalker98a, Lauchli06, Wu08,  Yamaura12, Mourigal12, Janson16, Nilsen17, Orlova17, YanShannon17} and spin liquid phases~\cite{Bergman07, Balents10, Gardner10, Henley10, Castelnovo12, ZhouNg17, Trebst17}.
However, to distinguish these two types of phases is often tricky, since both of them are invisible to conventional magnetic measurements.
Indeed, there have been steady reports of ``hidden'' multipolar orders from a magnetically disordering state~\cite{Chalker92, Ritchey93, Reimers93, Zhitomirsky02, Zhitomirsky08, MomoiShannon06, Shannon06, Mydosh11, Paddison15, LiChen16, Takatsu16, Luo17, TaillefumierShannon17}.
Moreover, identifying the right characterization of a spin-nematic order can also be a nontrivial task.
For instance, in the low temperature phase of the classical Heisenberg-Kagom{\' e} antiferromagnet, a hidden quadrupolar order was found first~\cite{Chalker92}, followed by the realization of an additional octupolar order~\cite{Ritchey93} and its optimal order parameter~\cite{Zhitomirsky02, Zhitomirsky08}.

The aforementioned multipolar orders are only the simplest ones admitted by the subgroup structure of $O(3)$. There are indeed myriads of more complicated multipolar orders where even the abstract classification of their order parameters has only been accomplished {\it two} years ago~\cite{Nissinen16, Fel95, Haji-Akbari15}.
Along with the diverse interactions and lattice geometries in frustrated systems, identifying or ruling out certain orders becomes a difficult task for traditional methods, as there is no general rule to anticipate their presence or type.
Machine Learning then promises to cover a broad class of tentative orders without such prior knowledge.
Further, if the machine is interpretable, its result, e.g. the order parameter, can even be used as input to traditional methods.

In this Rapid Communication, we present a kernel method to probe general classical $O(3)$-breaking multipolar orders, and implement it by interpretable  support vector machines (SVMs)~\cite{PonteMelko17}.
We demonstrate its capacity by detecting various emergent multipolar orders up to rank $6$ and extracting their {\it analytical} order parameters.
In comparison to other machine learning schemes, such as neural networks, our method is {\it strong} interpretable and stable against diverse data sets and varying parameters.

{\it Model and samples.} We generate the training and testing samples by a gauge theory which can effectively simulate all possible $O(3)$-breaking multipolar orders.
The theory is defined by the Hamiltonian~\cite{Liu16}
\begin{align} \label{eq:gauge_model}
	H = \sum_{\langle i,j \rangle} \sum_{\alpha\beta\gamma} \mbb{J}^{\alpha \beta} \mathbf{S}^{\alpha}_i \cdot U^{\beta \gamma}_{ij} \mathbf{S}^{\gamma}_j,
\end{align}
on a cubic lattice.
At each lattice site $i$, there are three $O(3)$ spins, $\mathbf{S}^{\alpha}_i =
(S_{i,x}^{\alpha}, S_{i,y}^{\alpha}, S_{i,z}^{\alpha} ) $, labeled by a
`color' index $\alpha \in \{\mr{l}, \mr{m}, \mr{n} \}$.
These spins form local orthogonal triads and can represent
general spin rotations.
In addition, at each bond $\langle i,j \rangle$, there is a matrix gauge field,
$U_{ij}$, which takes values from a three-dimensional point group $G$ and mediates the interaction between neighboring spins, $U_{ij}
\in G \subset O(3)$.
 $\mathbb{J}$ is a coupling matrix. (See SM~\cite{SM} for more details.)

The gauge fields in Eq.~\eqref{eq:gauge_model} are used to control the nature of an emergent multipolar order.
For example, by choosing $G = D_{\infty h}$ or $G = D_{3}$, it will respectively recover the quadrupolar and (in-plane) octupolar
order with the same order parameter as their realization on a kagom{\'e} or triangular lattice~\cite{Zhitomirsky02, Zhitomirsky08, MomoiShannon06}.
This gives us the flexibility to validate our method against diverse and complicated multipolar orders.

The input data to the SVM are raw spin configurations $\mb{x} = \{ S^{\alpha}_{i, a} \}$.
We prepare these configurations by performing classical Monte Carlo simulations on Eq.~\eqref{eq:gauge_model}, even though the origin of the data is in principle arbitrary.
The simulations typically have been performed on lattices with volume $V = 16^3$, with about $0.4 \sim 4 \times 10^5$ total samples.

{\it SVM for multipolar orders.} The detection of a potential multipolar order is formulated as a supervised binary classification.
First, we collect a set of raw configurations $\{\mb{x}^{(k)}\}$ at temperatures $T^{(k)}$, serving as the training data, and assign each configuration a binary label, $y^{(k)} = \pm 1$.
These labels correspond to a disordered and an ordered class, and are determined comparing $T^{(k)}$ to a discriminatory temperature $T_\textup{disc}$.
$T_\textup{disc}$ does not need to coincide with the critical temperature $T_c$ since the SVM is robust against misclassified data.

Then the sequential minimal optimization algorithm~\cite{Platt98} is used to solve the underlying quadratic programming problem.
Consequently, we gain access to a decision function which predicts the label $y$ of a new sample $\mb{x}$.
The decision function is formally defined as
\begin{align}  \label{eq:d_formal}
  d(\mb{x}) &= \sum_{k} \lambda_k y_k K(\mb{x}^{(k)}, \mb{x}), & y &= \sgn(d(\mb x)).
\end{align}
$K(\mb{x}^{(k)}, \mb{x})$ is a kernel function which maps
the raw data to an auxiliary space where the data are separable by a hyperplane.
$\lambda_k$ are essentially Lagrange multipliers and are learnt during training.
They can be understood as the weight of a training sample $\mb{x}^{(k)}$ entering in the definition of the separating hyperplane (samples with $\lambda \neq 0$ are referred to as support vectors).

As realized in Ref.~\cite{PonteMelko17}, in addition to serving as a binary classifier, the decision function may also be regarded as some physical observable, given its form and meaning constraint by the choice of the kernel function.
Nonetheless, standard SVM kernels are introduced mainly for general applications in computer science, such as image classifications, and may not be optimal for physical systems.
Instead, one may consider introducing kernels that are designed with specific physical problems in mind.

This is particularly suitable for the probing of a potential multipolar order.
Mathematically, a multipolar order can generally be described by a tensor.
This allows us to define a kernel which is sensitive to general $O(3)$-breaking multipolar order,
\begin{align} \label{eq:kernel}
	K\big(\mb{x}^{\prime}, \mb{x}\big) &=  \big[\bds{\phi}(\mb{x}^{\prime}) \cdot \bds{\phi}(\mb{x})\big]^2, \\
	\mb{x} = \{ S^{\alpha}_{i,a} \} \ \mapsto \ \bds{\phi}(\mb{x}) &=\{\phi_{\mu}\}= \{ \langle S^{\alpha_1}_{a_1} \dots S^{\alpha_n}_{a_n} \rangle_{cl} \}.\label{eq:mapping}
\end{align}
Here $\bds{\phi}(\mb{x})$ is an (explicit) mapping that maps the raw spin configuration to monomials of degree $n$.
$\langle \dots\rangle_{cl}$ denotes a lattice average performed up to a finite spin cluster.
$\mu$ corresponds to collective indices $\mu = (\alpha_1,\dots,\alpha_n; a_1,\dots,a_n )$, $a_1, \dots, a_n \in \{x, y, z \}$, and $\alpha_1, \dots, \alpha_n$ run over spins with a cluster.
The spin cluster is used simply to reduce the computational complexity.
It is based on the property that a local order can be defined by a finite number of local fields.
(We will discuss the choice of the spin cluster below.)

With this kernel, the decision function Eq.\eqref{eq:d_formal} can be expressed as
\begin{align} \label{eq:decision}
	& d(\mb{x}) = \sum_{k} \lambda_k y_k \big[\bds{\phi} (\mb{x}^{(k)}) \cdot \bds{\phi}(\mb{x})\big]^2 = \sum_{\mu \nu} C_{\mu \nu} \phi_{\mu} \phi_{\nu}, \\
	& C_{\mu \nu} = \sum_{k} \lambda_k y_k \langle
	S^{\alpha_1}_{a_1} \dots S^{\alpha_n}_{a_n}\rangle_{cl}^{(k)} \langle S^{\alpha^{\prime}_1}_{a^{\prime}_1} \dots S^{\alpha^{\prime}_n}_{a^{\prime}_n} \rangle_{cl}^{(k)}.
\end{align}
$C_{\mu \nu}$ denotes a coefficient matrix constructed by support vectors and their weights, from which we can either identify an order and its {\it analytical} order parameter, or exclude the existence of an order.

To that end, the problem now lies in finding the explicit coordinates $c_{\bds{\alpha}}$ of a tensor $\mathbb{O}$ in a space $\mathcal{V}$ spanned by tensor bases of rank $n$,
\begin{align} \label{eq:O_tensor}
  \mathbb{O} = \sum_{\bds{\alpha}} c_{\bds{\alpha}} \mb{S}^{\alpha_1} \otimes \mb{S}^{\alpha_2} \otimes. .. \otimes \mb{S}^{\alpha_n}.
\end{align}
As a tentative order parameter of a multipolar order with the ground state manifold $O(3)/G$, this tensor needs to be invariant under the point group $G$.
Thus the relevant bases and their coefficients need to be correctly identified.
If an order is detected, $C_{\mu \nu}$ develops a ``pattern'' where this information can be systematically inferred.
Otherwise, it exhibits seemingly random noise from overfitting.

In other words, $C_{\mu \nu}$ defines the contraction of $\phi_{\mu}$ and $\phi_{\nu}$ in Eq.\eqref{eq:decision}. Consequently, the decision function can be related to the squared magnitude of the underlying order,
$d(\mb{x}) \sim \| \mathbb{O} \|_F^2 = \Tr \big(\mbb{O} \bds{\cdot} \mbb{O} \big)$,
where $\| \mathbb{O}\|_F$ is a tensor analogue of the Frobenius inner product.

The dimension of the tensor space depends on the number of spins, $r$, in the spin cluster as ${\rm dim}(\mathcal{V}) = r^n$.
A proper choice of the spin cluster will minimize the computational effort.
However, in cases where such a choice is not obvious, one can use a large spin cluster, and $C_{\mu \nu}$ will exhibit a periodic pattern from which the optimal cluster can then be inferred.
In the case of the gauge theory Eq.~\eqref{eq:gauge_model}, the three spins forming a local triad, $\{\mb{S}^l_i, \mb{S}^m_i, \mb{S}^n_i\}$, serve as such a cluster.

\begin{figure}
  \centering
  \includegraphics{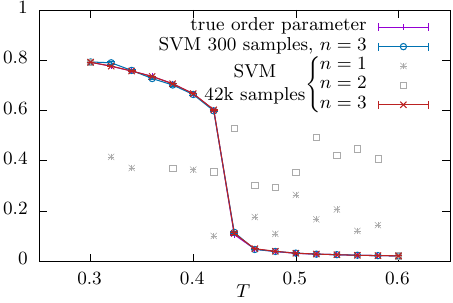}
  \caption{The square root of the decision function, $\sqrt{d(\mb{x})}$, trained
    at different ranks for the tetrahedral order. The true order parameter curve
    is shown for comparison, and $d(\mb{x})$ has been rescaled linearly, such
    that their endpoints match up. Insufficient tensor ranks do not result in a
    meaningful order parameter.}
  \label{fig:decisions}
\end{figure}

{\it High-rank orders.} We now apply the SVM equipped with the kernel~Eq.\eqref{eq:kernel} to probe an emergent tetrahedral ($T_d$), dodecahedral ($T_h$), octahedral ($O_h$) and icosahedral ($I_h$) order.
They represent the most complicated multipolar orders, going beyond the quadrupolar ($D_{\infty h}$) and the in-plane octupolar ($D_{3h}$) order.
It is important to emphasize that no prior knowledge about the existence and type of a potential multipolar order is required.

We start with the tetrahedral ($T_d$) order, training the SVM successively at the lowest ranks.
The discriminatory temperature $T_{\rm disc}$ is taken to be the ideal $T_c$ for now (cf. Ref.~\cite{Liu18}), but situations $T_{\rm disc} \neq T_c$ will be discussed later.

After training, the decision functions are measured for new testing samples.
The results are shown in Fig.~\ref{fig:decisions} by plotting $\sqrt{d(\mb{x})}$.
Clearly, the decision function exhibits only noise for lower ranks $n=1,2$, but converges at $n=3$, indicating that an order is captured at this rank.

\begin{figure}
  \centering
  \includegraphics{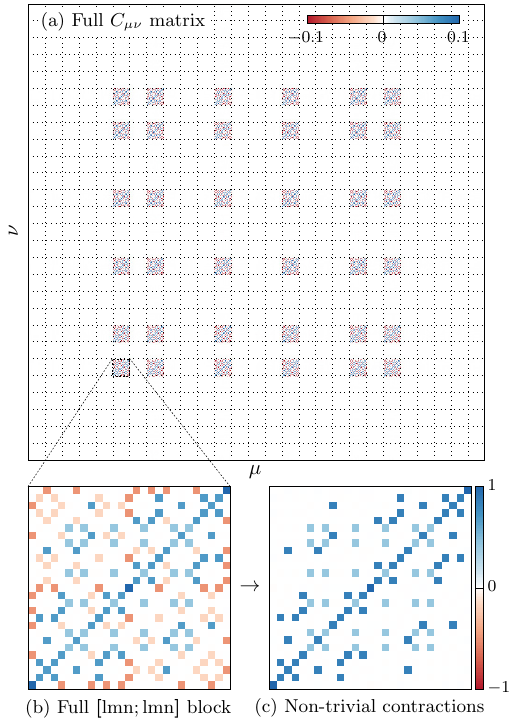}
  \caption{ The coefficient matrix $C_{\mu\nu}$ for the tetrahedral order learnt using the rank-$3$ kernel.
  (a) Full $C_{\mu\nu}$ matrix, where the multi-indices
  $\mu,\nu=(\alpha_1,\alpha_2,\alpha_3;a_1,a_2,a_3)$ are  lexicographically ordered.
  Each block is assigned  coordinates $\bds{[} \alpha_1 \alpha_2 \alpha_3; \alpha^{\prime}_1 \alpha^{\prime}_2 \alpha^{\prime}_3 \bds{]}$.
Nontrivial blocks have mutually exclusive color indices.
(b) Details of the ${\bds{[} \mr{lmn;lmn} \bds{]}}$ block, in comparison with (c) where trivial self-contractions have been removed.
  }
  \label{fig:Td}
\end{figure}

We then extract the order parameter from the corresponding $C_{\mu \nu}$ matrix.
At rank $3$, the general expression of the tensor $\mbb{O}$ in Eq.~\eqref{eq:O_tensor} involves $27$ basis tensors of the form
$\mb{T}^{\alpha_1\alpha_2\alpha_3} =
\mb{S}^{\alpha_1}\otimes\mb{S}^{\alpha_2}\otimes\mb{S}^{\alpha_3}$.
As shown in Fig.~\ref{fig:Td}(a), these divide $C_{\mu \nu}$ into $27$-by-$27$ blocks, and each block can be identified by their color indices as
 $\bds{[} \alpha_1 \alpha_2 \alpha_3; \alpha^{\prime}_1 \alpha^{\prime}_2 \alpha^{\prime}_3 \bds{]}$.
Only blocks with mutually exclusive color indices have nonvanishing entries. From this we can recognize the relevant basis tensors entering the definition of the underlying order parameter.
Furthermore, those blocks also exhibit an identical weight, by which the coefficients in $\mathbb{O}$ are also identified.
Thus the entire $C_{\mu\nu}$ matrix then corresponds to contracting two tensors,
 $\mathbb{O}^{(T_d)} = \sum_{\alpha_1 \neq \alpha_2 \neq \alpha_3} \mb{T}^{\alpha_1 \alpha_2 \alpha_3}$
which is exactly the tetrahedral order parameter \cite{Fel95}.
Consistently, the decision function is related to its norm squared,
$d(\mb{x}) \sim \| \mathbb{O}^{(T_d)}\|_F^2$,
up to linear rescaling.

We now zoom into the details of a nontrivial block, e.g., the ${\bds{[} \mr{lmn;lmn} \bds{]}}$ shown in Fig.~\ref{fig:Td}(b).
Its entries correspond to all possible contractions of two basis tensors
$\mb{T}^{\alpha_1\alpha_2\alpha_3}$
and $\mb{T}^{\alpha^{\prime}_1\alpha^{\prime}_2\alpha^{\prime}_3}$.
These include proper contractions such as
 $T^{\alpha_1\alpha_2\alpha_3}_{a_1 a_2 a_3} T^{\alpha^{\prime}_1 \alpha^{\prime}_2 \alpha^{\prime}_3}_{a_3 a_2 a_1}$
 and self-contractions such as
  $T^{\alpha_1\alpha_2\alpha_3}_{a_1 a_1 a_3} T^{\alpha^{\prime}_1 \alpha^{\prime}_2 \alpha^{\prime}_3}_{a_3 a_2 a_2}$ which contract at least one pair of indices on the same tensor.
 The former type is consistent with the Frobenius inner product
${\Tr}\big(\mb{T}^{\alpha_1\alpha_2\alpha_3} \bds{\cdot} \mb{T}^{\alpha^{\prime}_1\alpha^{\prime}_2\alpha^{\prime}_3}\big)$,
and has nontrivial contributions to the decision function.
In contrast, the self-contractions only contribute a trivial constant to the decision function and can be systematically identified and removed by a least-squares fit, as shown in Fig.~\ref{fig:Td}(c).

The key insight here is that, if a multipolar order is detected, its order parameter can be inferred from the ``coordinates'' of nontrivial blocks and their relative weights, regardless of the details within each block.
Hence, the interpretation of $C_{\mu\nu}$ is rather straightforward.

\begin{figure}
  \centering
    \includegraphics{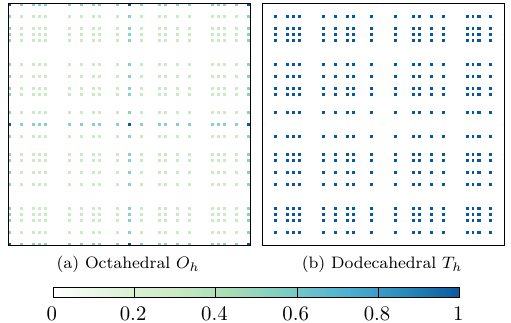}
  \caption{Block structure of $C_{\mu\nu}$ for (a) the octahedral and (b) the
    dodecahedral order, learnt using the rank-4 kernel. Each pixel corresponds
    to a block of $C_{\mu\nu}$, identified by coordinates
    $\bds{[}\alpha_1\alpha_2\alpha_3\alpha_4;\alpha'_1\alpha'_2\alpha'_3\alpha'_4\bds{]}$.
    The value of each pixel is given by the squared Frobenius norm of the
    corresponding block.}
  \label{fig:Oh_Th}
\end{figure}

This also holds true for the more complicated orders.
In Fig.~\ref{fig:Oh_Th}, we show such block structures of $C_{\mu \nu}$ for the octahedral ($O_h$) and dodecahedral ($T_h$).
For both cases, the order is learnt at rank $4$, and each block (pixel) is again identified by the spin color indices $\bds{[}\alpha_1\alpha_2\alpha_3\alpha_4;\alpha'_1\alpha'_2\alpha'_3\alpha'_4\bds{]}$.
The coordinates of the dominant blocks featuring four identical color indices ($O_h$) and two mutually exclusive pairs of identical color indices
($T_h$), respectively.
Correspondingly, their interpretations give rise to the ordering tensors,
$\mbb{O}^{(O_h)} = \mb{T}^{\mr{llll}} + \mb{T}^{\mr{mmmm}} + \mb{T}^{\mr{nnnn}}$
and
$\mbb{O}^{(T_h)} = \mb{T}^{\mr{llmm}} + \mb{T}^{\mr{mmnn}} + \mb{T}^{\mr{nnll}}$~\cite{Nissinen16}.
In particular, $\mbb{O}^{(T_h)}$ is a partially symmetric tensor and has six equivalent definitions generated by permuting its color indices.
Interestingly, SVM captures all these variants exhaustively.
Moreover, the subdominant blocks in Fig.~\ref{fig:Oh_Th} effectively remove the trace of $\mbb{O}^{(O_h)}$ and $\mbb{O}^{(T_h)}$, which does not change the decision function but is desirable in terms of SVM's optimization objective.
Such blocks do not occur in Fig.~\ref{fig:Td}a in the $T_d$ case as $\mbb{O}^{(T_d)}$ is traceless.

We also examined the icosahedral ($I_h$) order which is arguably the most complicated multipolar order breaking the $O(3)$ symmetry. We captured this order with a rank-$6$ kernel and extracted the rank-$6$ ordering tensor, $\mbb{O}^{(I_h)}$, from the block structure of the learnt $C_{\mu\nu}$,
$\mbb{O}^{(I_h)} = \sum_{\text{cyc}} \Big [ \bigl.\mathbf{S}^{\mathrm{l}}\bigr.^{\otimes 6} +\sum_{ \{+,- \}}
			\big(\frac{1}{2} \mathbf{S}^{\mathrm{l}} \pm \frac{\varphi}{2} \mathbf{S}^{\mathrm{m}} \pm \frac{1}{2\varphi} \mathbf{S}^{\mathrm{n}} \big)^{\otimes 6} \Big]$
where $\sum_{\text{cyc}}$ runs over cyclic permutations of three color indices.
This coincides with the exact result~\cite{Nilsen17} in which $\varphi = \frac{\sqrt{5}+1}{2} \approx 1.61803$ is the golden ratio. In comparison, we extracted a value of $\varphi = 1.61784$ (see the SM~\cite{SM} for details).

{\it Performance.} To quantify the performance of
SVM, we introduce a deviation metric, $\delta$, defined
by the element-wise discrepancy between the learnt $C_{\mu \nu}$ and the
theoretical one, $\tilde{C}_{\mu \nu}$, $\delta \coloneqq \frac{\|\mb{C} -
  \tilde{\mb{C}}\|_F}{\|\tilde{\mb{C}}\|_F} \geq 0$.
The tetrahedral order is taken as an example, but the general features are also
valid for the other aforementioned orders.

\begin{figure}
  \centering
  \includegraphics{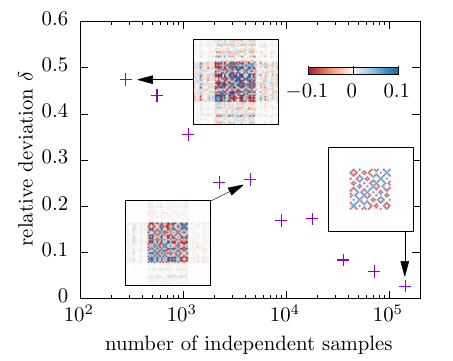}
  \caption{Deviation $\delta$ for the tetrahedral order as a function of the number of training samples.
  The insets show excerpts of the coefficient tensor for selected points.}
  \label{fig:samples}
\end{figure}

In Fig.~\ref{fig:samples} we demonstrate the dependence of $\delta$ on the number of
training samples. Interestingly, the expected block structure of $C_{\mu\nu}$ has
already emerged at as little as 300 samples, which is sufficient to infer the
underlying order parameter.
We emphasize that $\delta$ is a rather sensitive deviation metric. Empirically,
with a deviation $\delta \approx 0.5$, the measured decision function
$d(\mb{x})$ remains in decent agreement with the true order parameter curve
(Fig.~\ref{fig:decisions}).

\begin{figure}
  \centering
  \includegraphics{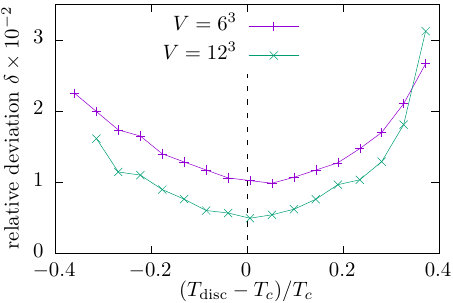}
  \caption{Deviation $\delta$ for the tetrahedral order against different
    discriminatory temperatures $T_\textup{disc}$ used to classify training
    samples.}
  \label{fig:Td_Tc}
\end{figure}

Figure~\ref{fig:Td_Tc} shows $\delta$ against the discrepancy of the assumed
$T_{\rm disc}$ from the real $T_c$.
We remark on the low level of error even for an estimate that is off by as much as
$ |\tau| \sim 40\% $, where $\tau = \frac{T_{\rm disc} - T_c}{T_c}$.
This robustness of the SVM facilitates applications where the locus of the phase
transition is not known {\it a priori}.
Moreover, a crude $C_{\mu \nu}$ learnt with large $|\tau|$ can in turn guide a
better estimate of $T_c$, as a well behaved $d(\mb{x})$ is still obtained,
reminiscent of the learning-by-confusion scheme \cite{Nieuwenburg17}.
Additionally, as seen from Fig.~\ref{fig:samples}, a crude $C_{\mu \nu}$ may
already suffice for an appropriate inference of the potential order parameter by
which one could further derive more sensitive measurements of a phase transition,
such as the susceptibility and Binder cumulant.

{\it Concluding remarks.} We have presented an interpretable kernel method to probe emergent multipolar orders and their {\it analytical} order parameter, which demands no prior knowledge about their existence.
We demonstrated its capabilities by addressing the most intricate representatives of these orders, and showed its stability against uncertainty regarding phase boundary and modest amounts of training data.
Our method can be used to detect the ``hidden'' multipolar orders in frustrated spin and orbital systems, especially when the presence of such orders is obscured by complex interactions and lattice structures.
Alternatively, it can also provide more exhaustive scrutiny towards spin liquid candidates, in comparison with conventional methods for excluding symmetry-breaking orders.
Furthermore, although we exclusively use SVMs in this work, our kernel can also be employed by other kernel methods such as kernel principle component analysis.

{\it Note added in proof.} In a followup work, we have further developed the method to cope with multiple (coexisting) orders and to explore unknown phase diagrams~\cite{svm2}.

\begin{acknowledgements}
This work is supported by FP7/ERC Consolidator Grant No. 771891, the Nanosystems Initiative Munich (NIM), and the Deutsche Forschungsgemeinschaft (DFG, German Research Foundation) under Germany's Excellence Strategy--EXC-2111--390814868.
We would like to thank Lei Wang for enlightening discussions.
Our simulations make use of the $\nu$-SVM formulation \cite{Scholkopf00}, the
LIBSVM library \cite{Chang01, Chang11}, and the ALPSCore library
\cite{Gaenko17}.
\nocite{Elitzur75}
The source codes and raw data supporting the findings of this study have been made
openly available \cite{Jonas}.
\end{acknowledgements}

\bibliographystyle{apsrev4-1}
\bibliography{svm}

\onecolumngrid
\clearpage
\makeatletter
\begin{center}
  \textbf{\large --- Supplementary Materials ---\\[0.5em]\@title}\\[1em]

  Jonas Greitemann, Ke Liu, and Lode Pollet
  \thispagestyle{titlepage}
\end{center}
\setcounter{equation}{0}
\setcounter{figure}{0}
\setcounter{table}{0}
\setcounter{page}{1}
\setcounter{section}{0}
\renewcommand{\theequation}{S\arabic{equation}}
\renewcommand{\thefigure}{S\arabic{figure}}
\renewcommand{\thetable}{S\arabic{table}}
\renewcommand{\thesection}{S.\Roman{section}}

\twocolumngrid

\section{The effective gauge theory}
The effective gauge theory, Eq.~\eqref{eq:gauge_model}, can be expressed in component form where Einstein summation is understood,
\begin{align}
H = \sum_{\langle i,j \rangle} J^{\alpha \beta}_{ab}S^{\alpha}_{i,a} U^{\beta \gamma}_{ij} S^{\gamma}_{j,b}.
\end{align}
It is reminiscent of the Hamiltonian of generalized exchange interaction of nearest-neighboring spins,
$H_{\rm ex} = \sum_{\langle i,j \rangle} J_{ab} S_{i,a} S_{j,b}$.
However, owing to the presence of gauge fields $U_{ij}$, Eq.~\eqref{eq:gauge_model} possesses a local symmetry defined by the gauge transformation
\begin{align}
  S^{\alpha}_{i,a} &\mapsto \Lambda^{\alpha \alpha^{\prime}}_i S^{\alpha^{\prime}}_{i,a},\\
  U^{\alpha \beta}_{ij} &\mapsto \Lambda^{\alpha \alpha^{\prime}}_i U^{\alpha^{\prime} \beta^{\prime}}_{ij} \Lambda^{\beta^{\prime} \beta}_j, &&
  \raisebox{.75\normalbaselineskip}[0pt][0pt]{$\forall\ \Lambda_i,\Lambda_j \in G.$}
\end{align}
Correspondingly, the global $O(3)$ symmetry of Eq.~\eqref{eq:gauge_model} is defined as
\begin{align}
	S^{\alpha}_{i,a} &\mapsto S^{\alpha}_{i,a^{\prime}} \Omega_{a^{\prime} a}, &&\forall\ \Omega \in O(3).
\end{align}

The general form of the coupling $J^{\alpha \beta}_{ab}$ is constrained by the gauge symmetry,
\begin{align}
	  J^{\alpha \beta}_{ab} & =
	   \Lambda^{\alpha \alpha^{\prime}}
	   J^{\alpha^{\prime} \beta^{\prime}}_{ab}
	 	\Lambda^{\beta^{\prime} \beta}, &&\forall\ \Lambda \in G.
\end{align}
However, to realize a multipolar order, it is sufficient to work with the isotropic limit $J^{\alpha
  \beta}_{ab} = J \delta_{\alpha \beta} \delta_{ab}$.
  The sign of $J$ is not critical as it may be absorbed by a gauge transformation.

As the gauge symmetry cannot break spontaneously~\cite{Elitzur75}, the model Eq.~\eqref{eq:gauge_model} develops multipolar order with ground state manifold $O(3)/G$.
Different types of multipolar orders can thus be realized by simply varying the gauge symmetry $G$ ~\cite{Liu16}.
This is convenient for us to generate training and testing samples to validate our method against various complicated orders.
However, the origin of these data is arbitrary to our method.

\section{Complexity and redundancy of the monomial mapping}

In order for the data $\{\mathbf{x}\}$ to become separable by a linear classifier such as SVM, one has to map it to a (typically) higher-dimensional feature space. By virtue of the \emph{kernel trick}, this mapping does not have to be known explicitly, as it is sufficient to be able to calculate inner products in that feature space. The kernel function $K(\mathbf{x'},\mathbf{x})$ then acts as a stand-in for the inner product.

The kernel we propose in the main text is a composition of the quadratic kernel $K_\text{quad}$ with the monomial mapping $\bds{\phi}$, cf. Eq.~\eqref{eq:mapping}:
\begin{align}
  K(\mathbf{x'},\mathbf{x}) = K_\text{quad}(\bds{\phi}(\mathbf{x'}),\bds{\phi}(\mathbf{x})) = \bigl[ \bds{\phi}(\mathbf{x'})\cdot\bds{\phi}(\mathbf{x}) \bigr]^2.
\end{align}
However, it turns out that the monomial mapping actually reduces the dimension of the feature space and it is therefore prudent to carry it out explicitly and to rely on the kernel trick only for the implicit mapping to an indeed higher-dimensional space due to the quadratic kernel.

In particular, by eliminating redundant elements, i.e. including the monomial
$
  \langle S_{a_1}^{\alpha_1} S_{a_2}^{\alpha_2} \dots S_{a_n}^{\alpha_n} \rangle_{cl}
$
in $\bds{\phi}(\mb{x})$ if and only if $(\alpha_1,a_1)\le
(\alpha_2,a_2)\le\dots\le(\alpha_n,a_n)$ with some arbitrary ordering imposed on
color-component tuples, the dimension of $\bds{\phi}(\mb{x})$ is given by the
multiset coefficient
\begin{align}
  \dim\bds{\phi}(\mb{x}) = \left(\!\!\!\binom{3r}{n}\!\!\!\right) = \binom{3r+n-1}{n} = \frac{(3r+n-1)!}{n!(3r-1)!}.
\end{align}
Table~\ref{tab:dims} explicitly demonstrates the growth of the configuration
vector with rank.

\begin{table}
  \centering \renewcommand{\arraystretch}{1.5}
  \setlength{\tabcolsep}{8pt}
  \begin{tabular}{r|*{6}{C}} \toprule
    rank $n$ & 1 & 2 & 3 & 4 & 5 & 6\\\hline
    $(3r)^n$ & 9 & 81 & 729 & 6561 & 59049 & 531441\\
    $\multiset{3r}{n}$ & 9 & 45 & 165 & 495 & 1287 & 3003\\ \bottomrule
  \end{tabular}
  \caption{Dimensions of the configuration vector $\bds{\phi}(\mb{x})$
    before and after eliminating redundant monomials. $3r=9$ is the range of the
    spin indices $(\alpha,a)$ ($r=3$ colors, 3 components).}
  \label{tab:dims}
\end{table}

By contrast, $\dim\mb{x}$ is extensive in volume and---in case of the gauge
model---will outgrow $\dim\bds{\phi}(\mb{x})$ for lattices as small as $L\ge 7$
even when the rank-6 mapping is used.

The explicit mapping of the configuration vector $\mb{x}$ to monomials thus provides a substantial advantage by removing the dependence on the lattice size, but comes
with the downside that $\dim\bds{\phi}(\mb{x})$ depends on the tensor rank $n$
which needs to be fixed when sampling configurations from the Monte Carlo
simulation.

For the interpretation of the coefficient tensor, we find it more
beneficial to include redundant elements to avoid obfuscating the block structure
discussed in the main text. The multiplicity, i.e. the number of equivalent
permutations, is given by the multinomial coefficients
\begin{align}
  m_{(\alpha_1,a_1)\dots(\alpha_n,a_n)} = \binom{n}{k_1,k_2,\dots} &= \frac{n!}{k_1!k_2!\dots},
\end{align}
where $k_1+k_2\dots+k_{3r}=n$ count the occurrences of each of the $3r$
possible index values. We include the square roots of the multiplicities
in the configuration, e.g. at rank 2,
\begin{align}
  \bds{\phi}(\mb{x}) = \{\sqrt{m_{\dots}}\mean{S_{a_1}^{\alpha_1}S_{a_2}^{\alpha_2}}_{cl}\ |\ (\alpha_1,a_1)\le(\alpha_2,a_2)\}.
\end{align}
That way, when using the above configuration in conjunction with the quadratic
kernel, we learn the same decision function that one would have gotten if all
$(3r)^n$ monomials had been considered regardless of their redundancy.

\section{Regularization parameter}

SVMs involve a regularization parameter $C$ and its choice, as it applies to phase
classification, has been discussed previously \cite{PonteMelko17}. In principle,
one has to validate the learnt model with respect to independent test data for
different values of $C$ which can span many orders of magnitude.

There exists however an alternative reparametrization of the SVM optimization
problem in terms of a regularization parameter $\nu\in[0,1)$ which has been
shown to impose a lower bound on the fraction of training samples that serve as
support vectors \cite{Scholkopf00}. $\nu$-SVM thus admits a more universal
interpretation and we found it to simplify the selection of an appropriate
regularization.

\begin{figure}
  \centering
  \includegraphics[scale=0.97]{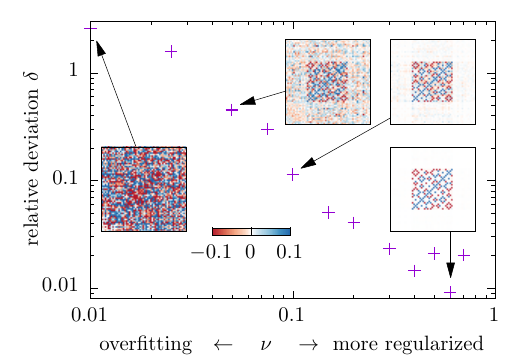}
  \caption{Deviation $\delta$ for the tetrahedral order, computed with different
    levels of regularization $\nu$.}
  \label{fig:nu}
\end{figure}

For the present work, we found a stronger regularization in terms of $\nu$ to
improve the quality of the learnt order parameter as demonstrated in
Fig.~\ref{fig:nu} for the tetrahedral order. This is consistent with the
fact the ensembles of micro-states in either phase near the transition
temperature have a significant overlap. Thus, we picked a rather large value of
$\nu=0.6$ for the data presented in Figs.~\ref{fig:decisions}-\ref{fig:samples},
and $\nu=0.4$ in Fig.~\ref{fig:Td_Tc} to allow for more imbalanced training data
\cite{Chang11}.

\

\section{Icosahedral order}
The icosahedral ($I_h$) order is captured by SVM with a rank-$6$ kernel, and we extract a coefficient matrix whose elements are denoted by
$C_{\mu \nu} = C^{\bds{\alpha}\bds{\beta}}_{\mb{a}\mb{b}} = C^{\alpha_1 \dots \alpha_6, \beta_1 \dots \beta_6}_{a_1 \dots a_6, b_1 \dots b_6}$.
This coefficient matrix is divided into $729$-by-$729$ blocks, identified by their color indices
$[\bds{\alpha},\bds{\beta}]$.
The underlying order parameter tensor is again inferred from the coordinates and the relative weight between non-vanishing blocks.
There are different ways to define the weight of a block.
For example, one can perform a summation over the the spin component indices for a given block $[\bds{\alpha},\bds{\beta}]$ as
$B^{\bds{\alpha\beta}} = \sum_{\mb{a}\mb{b}} C^{\bds{\alpha}\bds{\beta}}_{\mb{a}\mb{b}}$,
where $B^{\bds{\alpha\beta}}$ forms a reduced coefficient matrix.
Alternatively, one can also define the weight by the Frobenius norm of the corresponding block,
 $F^{\bds{\alpha\beta}} = \sqrt{\sum_{\mb{ab}} (C^{\bds{\alpha}\bds{\beta}}_{\mb{a}\mb{b}})^2}$.
 We have adopted both definitions and verified that the interpretations of the resulting reduced matrices lead to the same tensor.
 Moreover, we note that the reduced coefficient matrices are only used to facilitate the interpretation.

In terms of the reduced coefficient matrix, namely the block structure of the $C_{\mu \nu}$  matrix, we extract the icosahedral order parameter, $\mbb{O}^{(I_h)}_{S\rm VM}$, in the form of
\begin{align} \label{eq:Ih_SVM}
	\mbb{O}^{(I_h)}_{\rm SVM} &= \ 0.48657 \ \bigl.\mb{S}^{\mathrm{l}}\bigr.^{\otimes 6} - \ \bigl.\mb{S}^{\mathrm{l}}\bigr.^{\otimes 4} \bigl.\mb{S}^{\mathrm{m}}\bigr.^{\otimes 2} \nonumber + 0.51608 \ \bigl.\mb{S}^{\mathrm{l}}\bigr.^{\otimes 4} \bigl.\mb{S}^{\mathrm{n}}\bigr.^{\otimes 2} \nonumber \\
	&\quad + 0.48450 \ \bigl.\mb{S}^{\mathrm{l}}\bigr.^{\otimes 2}\bigl.\mb{S}^{\mathrm{m}}\bigr.^{\otimes 2}\bigl.\mb{S}^{\mathrm{n}}\bigr.^{\otimes 2} + \dots,
\end{align}
which contains $183$ terms in total, including all the combinations where each color index occurs an even number of times. (See the supplementary data for the full tensor.)
The coefficients of these terms are normalized against the $\bigl.\mb{S}^{\mathrm{l}}\bigr.^{\otimes 4} \bigl.\mb{S}^{\mathrm{m}}\bigr.^{\otimes 2}$ term, but this choice is arbitrary.
For comparison, the exact icosahedral ordering tensor, $\mbb{O}^{(I_h)}_{\rm ext}$, is defined as \cite{Nissinen16}
\begin{widetext}
\begin{align} \label{eq:Ih_ext}
	\mathbb{O}^{I_h}_{\rm ext} &= \sum_{\rm{cyc}} \left[ \bigl.\mb{S}^{\mathrm{l}}\bigr.^{\otimes 6} +\sum_{ \{+,- \}}
			\left(\frac{1}{2} \mb{S}^{\mathrm{l}} \pm \frac{\varphi}{2} \mb{S}^{\mathrm{m}} \pm \frac{1}{2\varphi} \mb{S}^{\mathrm{n}} \right)^{\otimes 6} \right]
				-\frac{1}{7} \sum_{\rm comb} \delta_{\alpha_1 \alpha_2} \delta_{\alpha_3 \alpha_4} \delta_{\alpha_5 \alpha_6}  \mb{S}^{\alpha_1} \otimes ...\otimes \mb{S}^{\alpha_6} \nonumber \\
  		&= \frac{7\varphi - 1}{112} \ \left(
  		\frac{5}{7\varphi-1} \ \bigl.\mb{S}^{\mathrm{l}}\bigr.^{\otimes 6} - \ \bigl.\mb{S}^{\mathrm{l}}\bigr.^{\otimes 4} \bigl.\mb{S}^{\mathrm{m}}\bigr.^{\otimes 2} + \frac{7\varphi -6}{7\varphi -1} \ \bigl.\mb{S}^{\mathrm{l}}\bigr.^{\otimes 4} \bigl.\mb{S}^{\mathrm{n}}\bigr.^{\otimes 2}
			 + \frac{5}{7\varphi -1} \ \bigl.\mb{S}^{\mathrm{l}}\bigr.^{\otimes 2}\bigl.\mb{S}^{\mathrm{m}}\bigr.^{\otimes 2}\bigl.\mb{S}^{\mathrm{n}}\bigr.^{\otimes 2} + \dots
  		\right),
\end{align}
\end{widetext}
where $\varphi = \frac{\sqrt{5}+1}{2}$ is the golden ratio and $\sum_{\rm cyc}$ sums over cyclic permutations of $\{\mb{S}^{\mathrm{l}}, \mb{S}^{\mathrm{m}}, \mb{S}^{\mathrm{n}} \}$.
The second term in the first line of the above equation is introduced to make $\mbb{O}^{(I_h)}_{\rm ext}$ traceless, and $\sum_{\rm comb}$ runs over all non-equivalent combinations of the color indices.
This only differs from the ordering tensor learnt by SVM in terms of a global normalization factor as indicated in the second line of Eq.~\eqref{eq:Ih_ext}.
Moreover, by solving equations of the relative ratios between terms, we obtain an approximate value of $1.61784$ for the golden ratio.

\end{document}